\documentclass{sf2a-conf2017}
\usepackage{graphicx}
\usepackage{hyperref}
\usepackage[]{natbib}  
\usepackage{epstopdf}

\def\BibTeX{{\rm B\kern-.05em{\sc i\kern-.025em b}\kern-.08em
    T\kern-.1667em\lower.7ex\hbox{E}\kern-.125emX}}
\bibpunct{(}{)}{;}{a}{}{,}  %%%%%%%%%%%%%  A&A bibliography style
\begin{document}

\TitreGlobal{SF2A 2017}

%%-----------------------------------------------------------------
%%      the top matter
%%

\title{The SPHERE Data Center: a reference for high contrast imaging processing}

\runningtitle{The SPHERE Data Center}

\author{Ph. Delorme}\address{Univ. Grenoble Alpes, CNRS, IPAG, F-38000 Grenoble, France}
\author{N. Meunier$^1$}
\author{D. Albert$^1$}
\author{E. Lagadec}\address{Universit\'e C\^ote d'Azur, Observatoire de la C\^ote d'Azur, CNRS, Lagrange, France}
\author{H. Le Coroller}\address{Aix Marseille Univ, CNRS, LAM, Marseille, France}
\author{R. Galicher}\address{LESIA, Observatoire de Paris, PSL Research University, CNRS, Sorbonne Unniversit\'es, UPMC Univ. Paris 06, Univ. Paris Diderot, Sorbonne Paris Cit\'e}
\author{D. Mouillet$^1$}
\author{A. Boccaletti$^4$}
\author{D. Mesa}\address{INAF - Osservatorio Astronomica di Padova, Vicolo dell'Osservatorio 5, I-32122 Padova, Italy}
\author{J.-C. Meunier$^2$}
\author{J.-L. Beuzit.$^1$}
\author{A.-M. Lagrange$^1$}
\author{G. Chauvin$^1$}
\author{A. Sapone$^3$}
\author{M. Langlois}\address{CRAL, UMR 5574, CNRS, Université Lyon 1, Saint Genis Laval Cedex, France}
\author{A.-L. Maire}\address{Max-Planck-Institut f\"ur Astronomie, K\"onigstuhl 17, 69117, Heidelberg, Germany}
\author{M. Montarg\`es}\address{Institute of Astronomy, KU Leuven, Celestijnenlaan 200D B2401, 3001 Leuven, Belgium}
\author{R. Gratton$^5$}
\author{A. Vigan$^3$}
\author{C. Surace$^3$}
\author{C Moreau$^3$}
\author{Th. Fenouillet$^3$}

%%%Institut de Radioastronomie Millim\'etrique, Saint Martin d’H\`eres, France

%% IF Author3 has the same affiliation than Author1:
%\author{C.\,E. Author3$^1$}

%% IF Author3 has its own affiliation:
%\author{C.\,E. Author3}\address{Dept. of Chess, University of Games, 35101 Las Vegas, Monaco} 

%% IF Author3 has two affiliations, the one of Author1 and a second one:
%\author{C.\,E. Author3$^{1,}$}\address{Dept. of Chess, University of Games, 35101 Las Vegas, Monaco} 

%% Keep this line, even if the page will be settled afterwards.
\setcounter{page}{237}

%%-----------------------------------------------------------------

\maketitle

%%-----------------------------------------------------------------
%%        The abstract
%% 
%%  Warning!  within the abstract:
%%  - do not use macros. 
%%  - do not use commands like: \cite, \citet, \citep ... etc.

\begin{abstract}
The objective of the SPHERE Data Center is to optimize the scientific return of SPHERE at the VLT, by providing optimized reduction procedures, services to users and publicly available reduced data. 
This paper describes our motivation, the implementation of the service (partners, infrastructure and developments), services, description of the on-line data, and future developments. 
The SPHERE Data Center is operational and has already provided reduced data with a good reactivity to many observers. The first public reduced data have been made available in 2017. The SPHERE Data Center is gathering a strong expertise on SPHERE data and is in a very good position to propose new reduced data in the future, as well as improved reduction procedures. 
\end{abstract}

%% Insert the keywords (to appear in the ADS indexing)
%% Keywords must be separated by a comma

%%-----------------------------------------------------------------

\section{Introduction: Objectives of the service}

\subsection{The SPHERE instrument}

The main goal of the SPHERE instrument \cite[Spectro-Polarimetric High-Contrast Exoplanet REsearch,][]{Beuzit.2008} on the VLT is to detect and 
characterize young giant exoplanets orbiting nearby stars using direct imaging. SPHERE also allows to detect and characterize circumstellar
disks where such exoplanets form. Other science cases include: stellar physics (massive stars, stars in clusters, stellar surface imaging), circumstellar environments (envelopes, jets, companions), planetology (asteroids, satellites), high energy objects, etc.  

The detection of exoplanets using direct imaging is very challenging because they are forming close to their parent star and their 
luminosity is much lower (typically by a factor of at least 10$^5$). The conception of SPHERE was therefore constrained by the necessity to obtain 
the largest contrast possible in the close environment of these young stars. As a consequence, SPHERE combines several 
sophisticated techniques to achieve this goal: extreme Adaptive Optics, coronagraphy, polarimetry, low resolution spectroscopy, differential imaging in the visible and near-IR and spectro-imaging in the near-IR. SPHERE is composed of three instruments fed by a common path: IRDIS (differential imaging, long-slit spectroscopy and polarimetry
in the near infra-red), IFS (integral field spectrograph, also in the near infra-red), and ZIMPOL (differential polarimetric imaging in the
visible). These instruments provide large quantities of heterogeneous data needing specific data processing to maximize the contrast
and more generally in order to derive the best performances in terms of detection and characterization of the astronomical signal. 

\subsection{Motivations : short- and long-term strategy}

Given the complexity of the instrument and of the data calibration specifically required to a variety of observing modes, it was necessary to implement a service that would allow to optimise the 
scientific return of the instrument, by helping the PI to process their data adequately, and by providing some state-of-art reduced data to the whole community.
In parallel, the way to optimally combine data and self-calibrate the stellar residual is still an active field of research, and new solutions are regularly proposed in the litterature: the development of dedicated tools and the build-up of a centralised expertise on massive Adaptive Optics imaging data reduction provide the community  with a support by experienced users who are likely to have already met the same issue and to have already developed a solution. Beside this scientific return optimisation, the database format containing massive amount of data and processing them consistently allows to develop in the near future unique complementary capabilities such as the coupling between wavefront sensor information and the actual detection limits on the final reduced images, the building of a massive reference library of SPHERE coronagraphic images for use in advanced reference differential imaging techniques as well as to explore the opportunity to use deep-learning techniques in AO analysis.

Our objective is to build-up a reference center on high-contrast imaging and gather the community to foster data reduction through centralised user-feedback and to continuously strengthen this community. On a long-term basis, such services could be extended to future instruments, in line with our ongoing service to provide reduced data for (older) Adaptive Optics instruments other than SPHERE (such as NACO and NICI for instance) in the DIVA archive (see Sect. \ref{DIVA}).

\section{Implementation of the service}

\subsection{Partners}

The SPHERE Data Center have been developed by four partners, with infrastructures localized on two sites: 
\begin{itemize}
\item{OSUG (Observatoire des Sciences de l'Univers de Grenoble)/ IPAG (Institut de Planetologie et d'Astrophy\-sique de Grenoble): The processing center (hereafter PC) is implemented in Grenoble. The most 
voluminous data (especially large data cubes) are available there, and this center also provides data reduction services. It is part of the OSUG Data Center\footnote{http://datacenter.osug.fr/}. }
\item{PYTHEAS (Observatoire des Sciences de l'Univers de Marseille) / LAM (Laboratore d'Astrophysique de Marseille): A database of 
high-level data, called the Target Data Base  (hereafter TDB) has been developed in Marseille, and receives data from the processing center in Grenoble, to which some complementary data are added. It is part of CeSAM\footnote{https://www.lam.fr/cesam/}. }
\item{OCA (Observatoire de la C\^ote d'Azur) / LAGRANGE laboratory: This partner contributes to the operation of the processing center. }
\item{Observatoire de Paris Meudon / LESIA (Laboratoire d'Etudes Spatiales et d'Instrumentation en Astrophysique): This partner contributes to the 
software development used by the processing center. }
\end{itemize}

Additional informal support was provided by members of the SPHERE Consortium who provided some of the algorithms used at the DC as well as contributed to debugging of the software. 

\subsection{Local infrastructures and developments}

   The SPHERE Data Center is organised around two main tools, the Processing Center (PC) and the Target DataBase (TDB). The PC is oriented toward data processing and reduced data distribution organised around observations and instruments while the TDB is oriented toward the distribution of advanced data products organised around targets and surveys. In the mid-term future, the TDB will be the main SPHERE Data Center link to the Virtual Observatory.

\subsubsection{The Processing Center }

\begin{figure}[ht!]
 \centering
 \includegraphics[width=1.0\textwidth,clip]{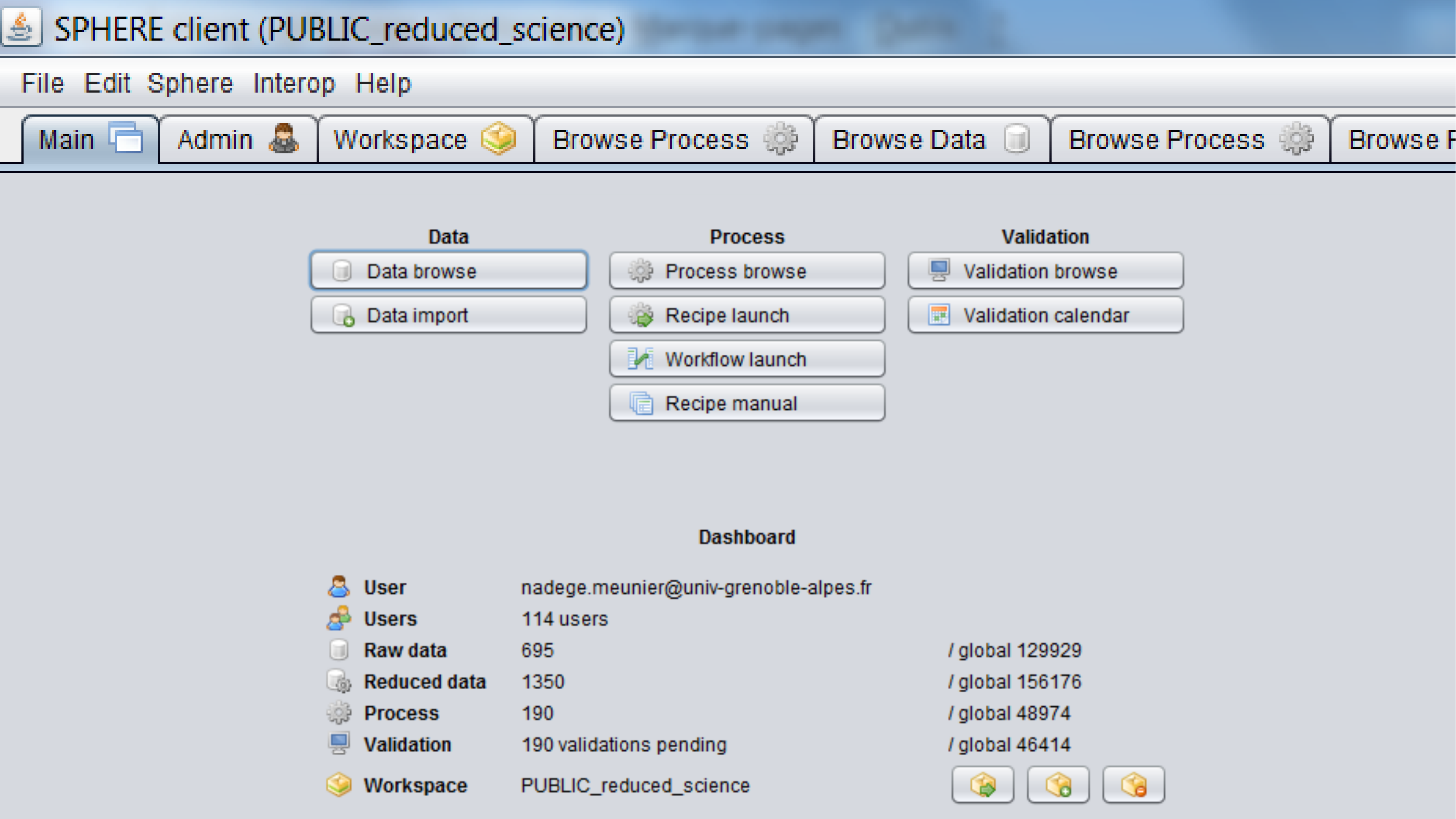}     
%%% Note the ABSENCE of the extension .pdf  !
  \caption{Example of the main menu on the Processing Center client interface (seen by administrators) and showing the different functionalities.}
  \label{fig1}
\end{figure}

\begin{figure}[ht!]
 \centering
 \includegraphics[width=1.0\textwidth,clip]{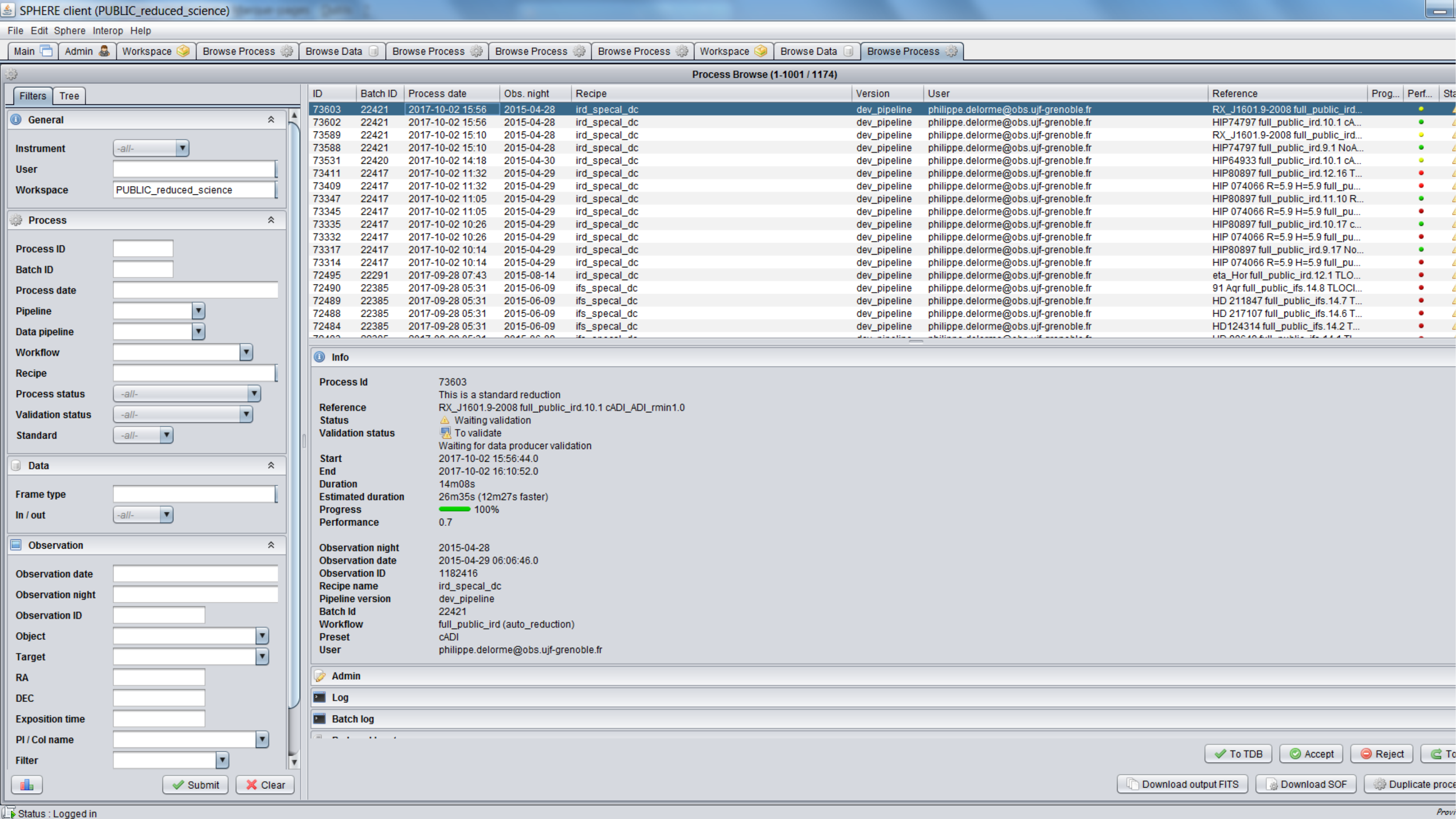}     
%%% Note the ABSENCE of the extension .pdf  !
  \caption{Example of the process browse functionality on the public reduced data, showing the main request criteria. }
  \label{fig2}
\end{figure}

The objective of the Processing Center was first to provide an infrastructure allowing a reduced team to easily process large amount of data in an homogeneous manner, and with a proper tracking system of the used data reduction parameters and algorithm versions. This data processing can be performed either on request with a good reactivity (since 2015), or systematically in a second phase of the project (started in 2017). It is therefore necessary to develop a 
complex database to archive many informations (and not only the information necessary for public data release) as well as a functionnal 
interface with many functionnalities. We summarize the different components developed at the Processing Center:

\begin{itemize}
\item{{\it A relational database}: The MySQL relational database contains raw and processed data and headers informations, processing parameters, 
processes which have been applied to the data, code version ... }
\item{{\it A user interface}: The interface allows the team to manage the operations: data import from ESO (manual after data retrieving, automatic for the calibrations) 
with file type identification, data browsing and processing, data validation, extraction of information for the instrument monitoring,
visualisation tools, ... This is illustrated in Fig.~\ref{fig1}. It also allows access to services and data (see fourth point below).  We used the jMCS framework developed by the JMMC (available at https://github.com/JMMC\_OpenDev/jMCS). }
\item{{\it The pipeline and workflows}: The pipeline is based on the DRH (Data Reduction and Handling) pipeline delivered by the SPHERE consortium to ESO \cite[][]{Pavlov.2008}, with a 
number of significant additions and corrections. It is indeed possible to add any new routine easily. The additionnal tools have been developed 
in Grenoble and Paris \cite[SpeCal pipeline,][]{Galicher.2017}  and are routinely used; more tools will be added in the future (see Sect.~5). The notion of workflows has been implemented, with standard workflows which are easy to use during processing operations: a significant effort has been dedicated to automatically associate the most relevant calibrations for each dataset. }
\item{{\it A user manual}: It is possible for external users to connect to the interface to work on their own data, using our application and hardware. Data is organized into workspaces for data protection and rights management. The interface therefore includes a user identification. A public username has been created for public data, which are located in specific workspaces, as illustrated in Fig.~\ref{fig2}. Additionnally, the SPHERE Data Center user manual is a valuable source of information for any user willing to reduce SPHERE data. }
\item{{\it Instrument monitoring}: Dedicated routines have been developed to extract useful information (Strehl ratio, seeing, coherence time, wave front sensor information) to monitor the instrument, based on the very large amount of data available. The SPHERE Data Center allows to quantitatively link such observational metadata to the final quality of the reduced products (see Sect.~5 for more details).}
\item{{\it Infrastructure}: Up to now a dedicated server was used for the software and data processing, associated to a data storage at Grenoble Alpes University (SUMMER). 
We are currently implementing a new server, dedicated to heavy data processing: this server will be part of a grid (LUKE) developed by the 
high performance computing mesocenter CIMENT at University Grenoble Alpes, with dedicated high performance storage disks. The SUMMER storage will continue to be used and will be extended over time, and an additional 
virtual server will then be implemented for the database at the OSUG Data Center (OSUG-DC) to manage the database and other operations.   }
\item{{\it Administration tools}: A number of functionnalities used to administrate the data center have been developed, such as (not exhaustive) user access and rights, logs, algorithm to recognize star names from SIMBAD at the CDS, monitoring tools (tasks progression, batchs, system health), reporting tools (data statictics, quality control reports), management tools (ESOrex pipeline versions, local DC recipes, workflows, users and security)}
\item{{\it Connection with the TDB}: The PC prepares advanced reduced data products gathering all necessary information that is automatically sent to the Target Data Base (TDB). }
\end{itemize}

\subsubsection{The Target Data Base (TDB)}

\label{tdb}

The Target Data Base, at http://cesam.lam.fr/spheretools/, has been developed in a very generic way, with the goal to provide access to high level data as well as dedicated tools to analyse them. 
In this section, we summarize the different components developed at the Target Data Base:\\

\begin{figure}[ht!]
 \centering
 \includegraphics[width=1.0\textwidth,clip]{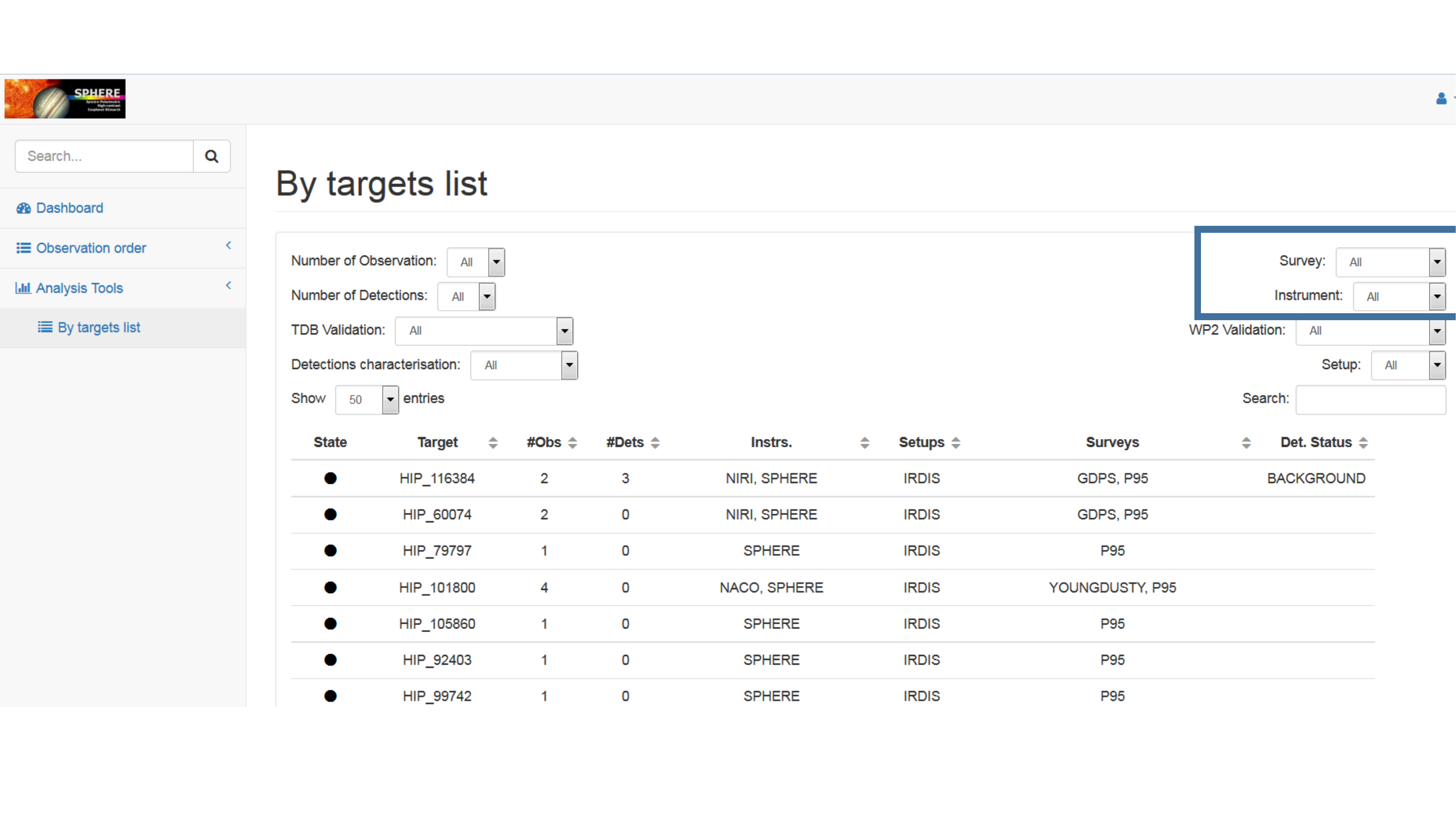}
%%% Note the ABSENCE of the extension .pdf  !
  \caption{Request form at the TDB. }
  \label{fig3}
\end{figure}

\begin{itemize}

\item{{\it A relational database}:
%\textbf{A relational data base} \\
The TDB is based on a hierarchical design  (Observations $->$ Reductions $->$ Detections\footnote{detections are objects (either companions or background stars) identified in the field of coronographic observations (see Sect.~5.2 for more details)}). 
A relational PostgreSQL database contains all the tables required
to store science products of high-contrast imaging surveys. The main tables and columns of this database are described in the Annexe \ref{tables}. 
In this structure, an observation is unique, and we can import as many reductions related to this observation as needed, as well as detections found in each reduction made by the PC.  
}
\item{{\it Data import and connexion to other databases}:
A generic code has been developed in the TDB  to import these data from a hierarchic json file\footnote{JavaScript Object Notation, file with an open-standard format allowing the storage and transfer of data structures and objects.}. This file format notably allows to import  data from various surveys even if some columns are missing (the surveys are not homogeneous, and sometime astrometry or photometry informations are missing). Such files are used to transfer data from the PC to the TDB (see Sect.~4 and 5 for the data description). 
%%%%Thus, already twelve public surveys (DIVA project\footnote{http://cesam.lam.fr/diva/} have been imported in the TDB.\\
%%%%All the SPHERE data reduced and pushed by the DC into the workspace TDB are automatically  imported via a web servcie.
The TDB also contains a lot of informations on the observed targets such as their astrometry (coordinates, proper motion, parallax), photometry (B, V, R, I, J , H, K magnitudes) and ages.
An algorithm of the TDB recognizes star names from SIMBAD at the CDS coupled to an internal dictionary for objects not referenced in SIMBAD, and astro-photometry of a new target is automatically retrieved from SIMBAD. 
%%%\noindent Important output files of the reductions are also transferred into the TDB via this web service (DC $->$ TDB).  These reduced data are:
}
%fin relational database
%\textbf{The web portal and tools}:\\
\item{ {\it The web portal and tools:}
The web portal (http://cesam.lam.fr/spheretools/) is mainly based on Python Django, and some pages use the ANIS framework \cite[][]{Moreau.2015} developed by the CeSAM in Marseille.
It allows requests on the database using forms (for example such as in Fig.~\ref{fig3}) with a large number of criteria and allowing crossing of informations: star age and mass, number of detections, confirmed companion candidates, reductions type, observation dates, etc. Request forms using the whole potential of the TDB relational data base will be developed in the future. In addition, tools such as dynamic charts have been developed for Level 3 data (see more details in Sect.~5.2).  
}
%fin web portal & tools
\item{{\it Rights management}:
A fine-grained rights management interface has been developed, based on Django. 
Depending on the rights of the users (public, in one or several surveys), it is possible to access different tools and/or data (see Sect.~5 for discussion).
}
\item{{\it Virtual Observatory}:
We plan to implement later Virtual Observatory service to give access to table data (service TAP=Table Access Protocol) and reference this service in the VO Registry. 
It will enable access to target and detection data by programs such as python scripts.
}
\end{itemize}

\section{Services}

\subsection{Services currently available}

\begin{itemize}
\item{{\it Data processing on request}: the Data Center team reduces PI data on demand and also carries out early data analysis steps, notably providing ADI analysis and detection limits. This service is  provided to the PI (or their coI) of the observation. \textit{For IRDIS, IFS and ZIMPOL}}
\item{{\it Access to the SPHERE Data Center application to process private data}: This service is an alternative to the first service. 
The user can access our application and process its own data through it. This service is open with some conditions: limited time 
for a given dataset (1 week), outside of intensive run periods (either GTO or systematic processing), so that we can control the charge on our server. This could evolve 
in the future depending on the needs and available infrastructures. \textit{For IRDIS and IFS}}
\item{{\it Support to SPHERE surveys}: Surveys or Large Programs may have specific needs. For example the SHINE survey of the SPHERE 
consortium has complementary needs in term of routines and organization, in order to organize data processing within H+24 of the actual observations. \textit{For IRDIS and IFS} }
\item{{\it Reduced calibration on-line}: Reduced SPHERE calibrations are put online less than two days after observation and can be easily downloaded with the public login. \textit{For IRDIS, IFS and ZIMPOL}}
\item{{\it Reduced observations on-line after a certain period}: Scientific SPHERE data are reduced, notably providing ADI analysis and associated detection limits. As of now most of P95 IFS and IRDIS science data is reduced and can be easily downloaded from the PC and TDB with the public login. Note that individual PIs have the possibility to delay the public release of reduced data obtained from their public raw data if this would endanger a publication in preparation and jeopardize their program, notably in case of the possible detection of a candidate companion at a first epoch, now public, for which the PIs are expecting a second epoch confirmation (this concerns a minority of observations). In the future we will strive to provide public reduced data of SPHERE science data one year after the raw data has been made public by ESO, so 2 years after the science observations in most cases. \textit{For IRDIS and IFS}}
\end{itemize}

%%%%%%\item{We gave individual PIs the possibility to delay the public release of reduced data obtained from their public raw data if this would endanger a publication in preparation, notably in case of the possible detection of a candidate companion at a first epoch, now public, for which the PIs are expecting a second epoch confirmation. This is in line with the objective of the SPHERE Data Center to provide a service to SPHERE users and not to jeopardize their program. Only a few PIs have chosen to delay the public release of their reduced data, which will be released later.  {\bf XXXX SE METTRE D'ACCORD SI on PARLE DE CA et si on en parle comme ça.XXXX}

\subsection{Practical information}

\begin{figure}[ht!]
 \centering
 \includegraphics[width=1.0\textwidth,clip]{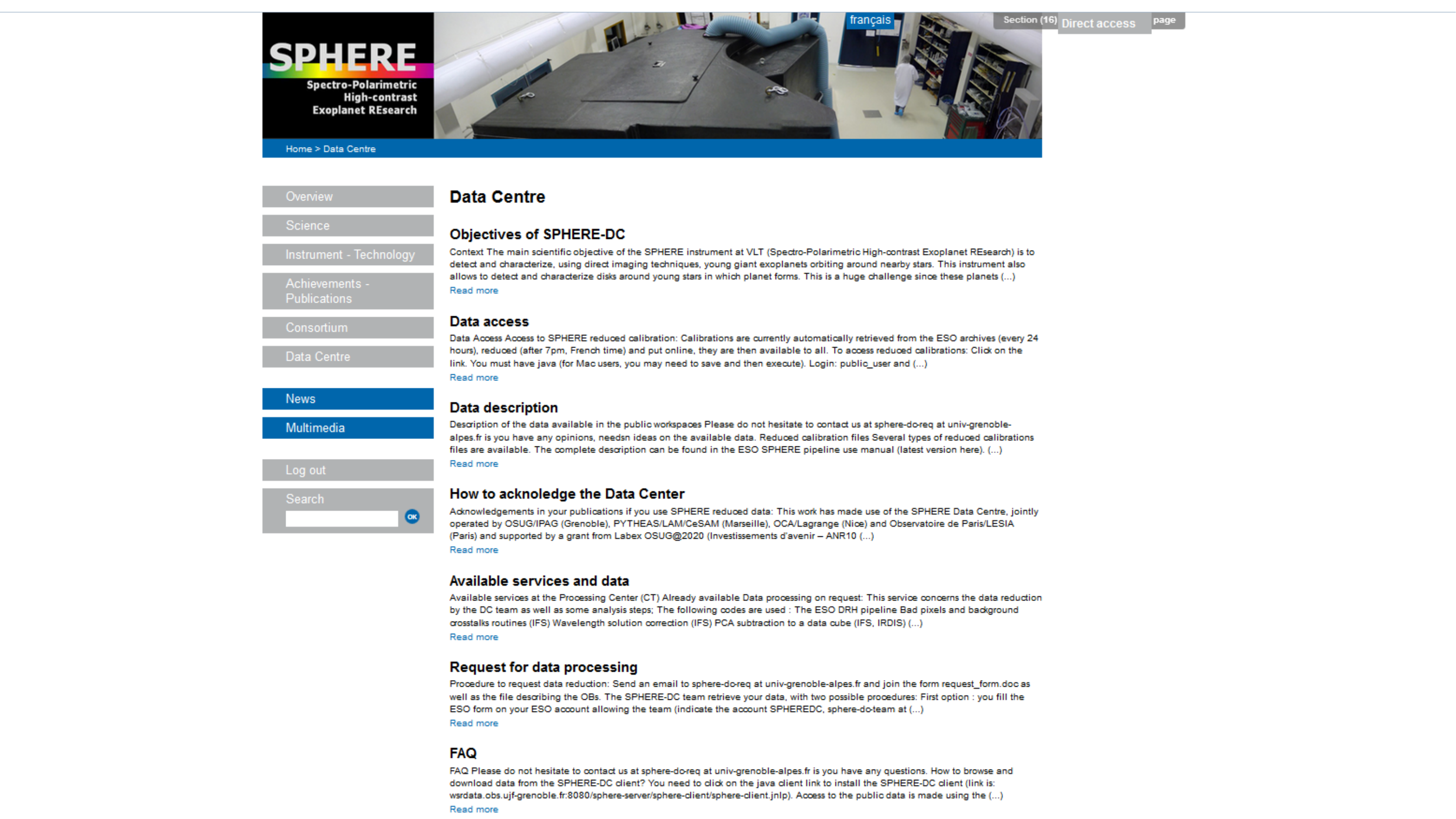}
%%% Note the ABSENCE of the extension .pdf  !
  \caption{Website of the SPHERE Data Center illustrating a subsample of the available information.}
  \label{fig4}
\end{figure}

Information on the SPHERE Data Center are available on our website at http://sphere.osug.fr/spip.php?rubrique\\16\&lang=en and will be regularly updated (Fig.~\ref{fig4}). They include (not exhaustive):

\begin{itemize}
\item{Information on data access, including a direct link to the TDBi\footnote{http://cesam.lam.fr/spheretools/search} and to DIVA\footnote{http://cesam.lam.fr/diva/}. }
\item{A FAQ with useful informations and tips.}
\item{A description of the available data.}
\end{itemize}

\section{On-line data: description of the data levels}

\subsection{Reduced calibrations}

In 2017 we have implemented an automatic retrieval procedure (from the ESO archive) to process all calibration files. They are retrieved 
every 24 hours. The reduction automatically starts at 7pm CET. These reduced 
calibration files are then publicly available in the PC database. Older calibration files will shortly be retrieved and 
processed as well.

\subsection{Level 1 data}

For IRDIS data, the first reductions steps (dark, flat, and bad pixel correction) rely on the SPHERE Data Reduction and Handling (DRH) pipeline \cite[][]{Pavlov.2008} provided by the consortium to ESO. The star centring is obtained through observations with coronagraph with waffle mode (a small fixed waffle pattern is applied to the deformable mirror to create four unsaturated crosswise echoes of the PSF) or with unsaturated exposures without coronagraph. The quality of this centring is automatically controlled and if the centring is too bad (error larger than 0.7pixel or 8.6mas) the workflows use a homemade centring routine to try to improve it and keep the most accurate result. 

For IFS data the SPHERE Data Center complements the DRH pipeline with additional steps that improve the wavelength calibration, bad pixel correction, and the cross-talk correction \citep{Mesa.2015}. The astrometric calibration, computation of parallactic angles (see below) are also done using routines external to the DRH.

For both IRDIS and IFS data, we automatically associate the calibrations that are the nearest in time (usually within 12hr of the observations) and use the same coronagraph, filter, neutral density, readout mode and exposure time (when relevant) as the science data. To perform the astrometric calibration of the IRDIS and IFS  dataset on sky, we use both the fixed instrumental corrections (0$^\circ$ for IRDIS in field tracking mode, 135.99$^\circ$ in pupil tracking mode, and -100.48$^\circ$ for IFS) and the on-sky calibration by \citet{Maire.2016astrom} leading to a True North correction value of $-1.75\pm0.08^\circ$ and a plate scale of  12.255$\pm$0.009 milliarcseconds/pixel for IRDIS and 7.46$\pm$0.02 milliarcseconds/pixel for IFS.

The resulting master reduced data cube, that stores one reduced frame per Detector Integration Time in a given observing block, is associated with the following metadata:
\begin{itemize}
\item{A vector containing the accurate parallactic angle for each frame. The raw headers have this information but with a coarser sampling (two per NDIT), so this vector is a prerequisite for using ADI algorithms for observations in pupil tracking mode.}
\item{A vector containing the wavelength calibration for each channel (2 for IRDIS and 39 for IFS).}
\item{If flux calibration data (unsaturated exposures of the out-of-coronagraph target star) are available, the Data Center reduces them and provides them together with the science data, for use as flux calibration or as PSF models.}
\item{In the near future, a table will be added that provides the essential information from the wavefront sensor (seeing, coherence time, wind speed and image quality estimator: Strehl ratio) derived with a frame by frame time-sampling, within the limit of the temporal sampling of recorded wavelength sensor statistical data: up to now every 20s, currently in discussion with ESO to improve to 10s.}
\end{itemize}

\begin{figure}[ht!]
 \centering
 \includegraphics[width=1.0\textwidth,clip]{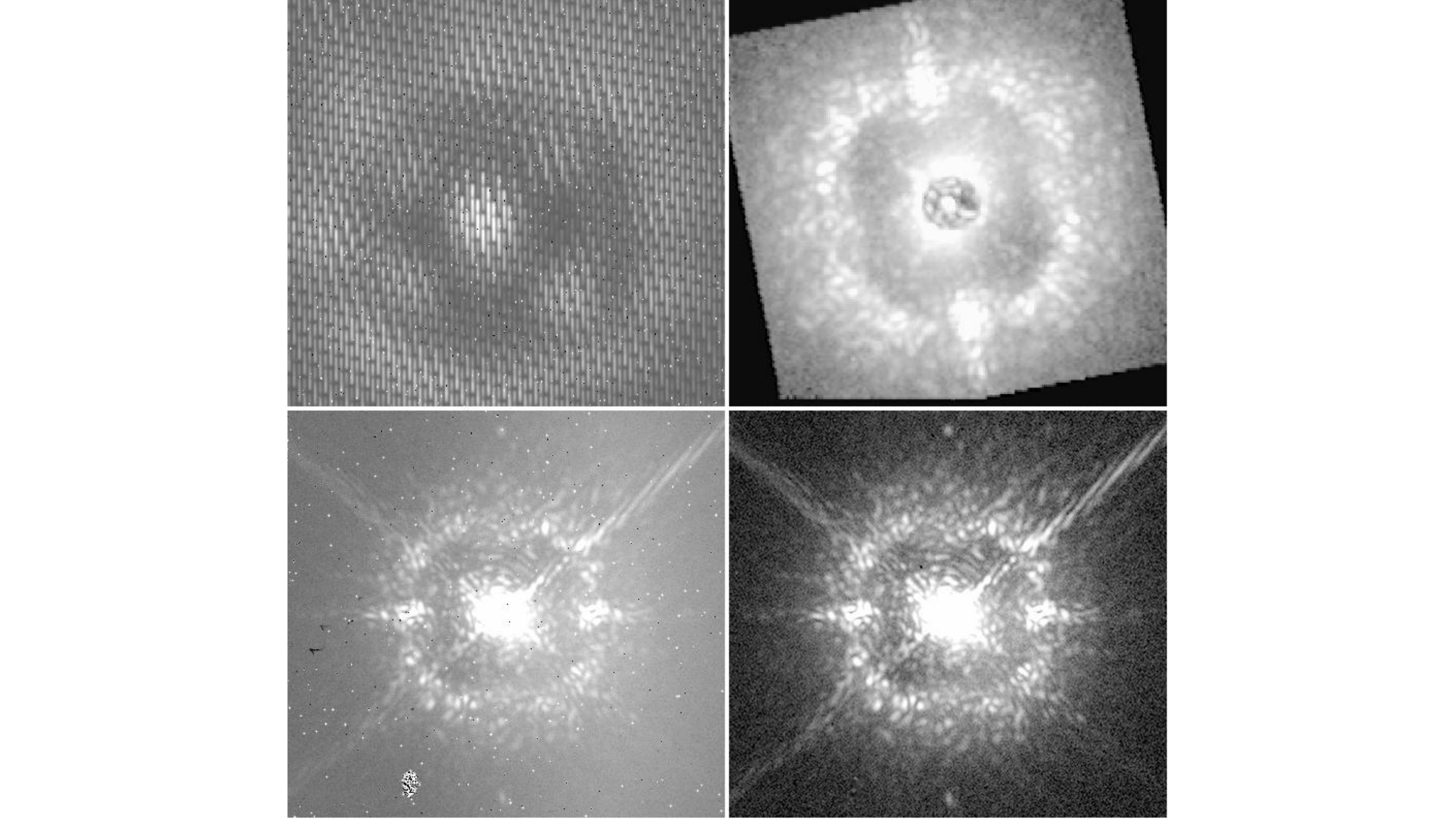}
%%% Note the ABSENCE of the extension .pdf  !
  \caption{Raw (left) and Level 1 reduced images (right) of HD206893B \cite[][]{Delorme.2017}, for IFS images (upper panels) and IRDIS (lower panels).
}
  \label{fig5}
\end{figure}

The resulting Level 1 reduced data are available through the PC, with the public login for public data or with private login for PI data. Examples of Level 1 images are shown on Fig.~\ref{fig5}.

\begin{figure}[ht!]
 \centering
 \includegraphics[width=1.0\textwidth,clip]{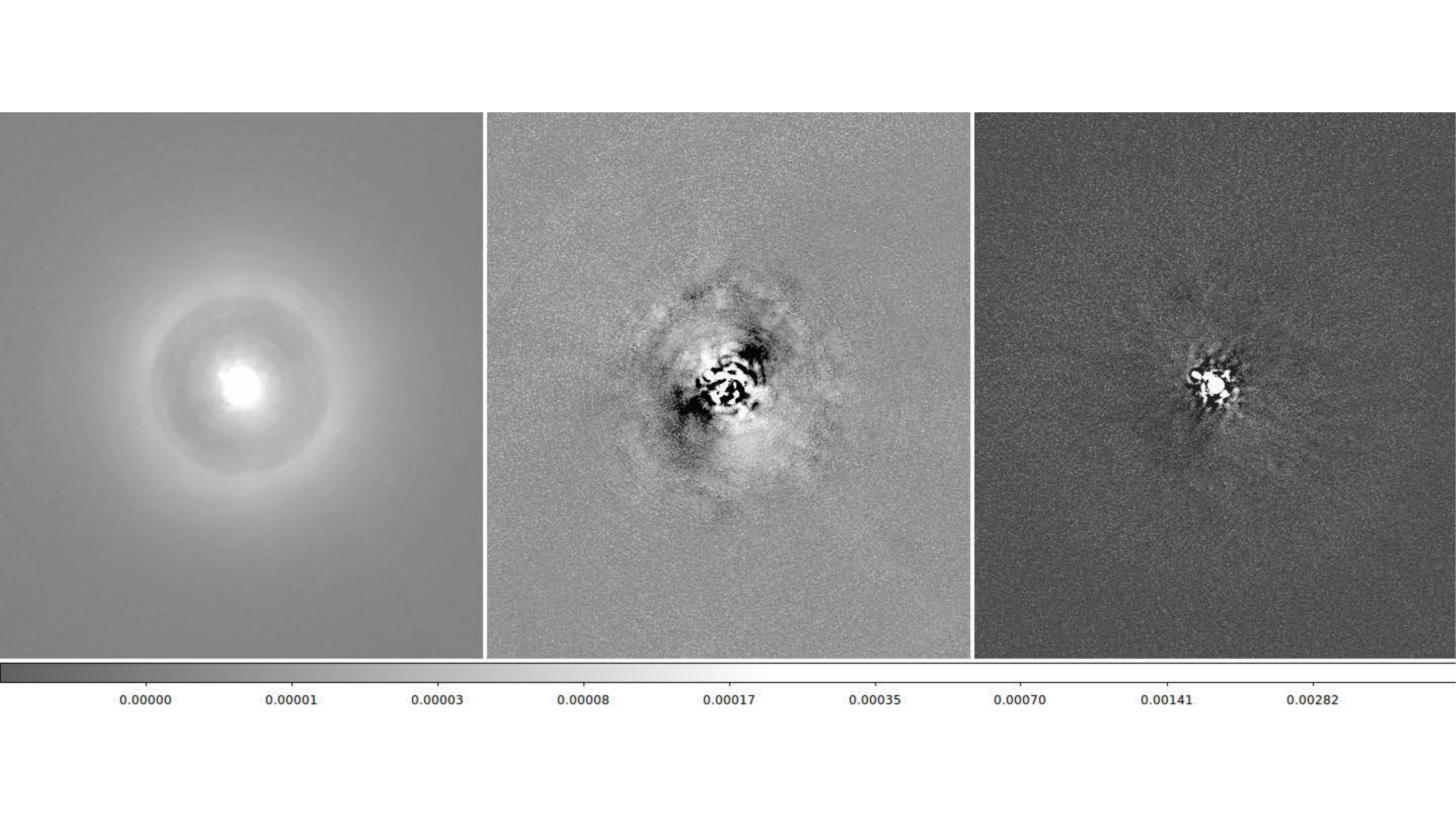}
%%% Note the ABSENCE of the extension .pdf  !
  \caption{Level 2 reduced residual of HD206893B \cite[][]{Delorme.2017} with various algorithms. From left to right: No ADI, Classical ADI, TLOCI-ADI. the substellar companion is visible on the upper left (North East) of the star center. }
  \label{fig6}
\end{figure}

\subsection{Level 2 data} \label{level2}

The Level 1 reduced master cubes are then used as input for high contrast imaging post-processing. We use the Speckle Calibration (SpeCal) package, described by \cite{Galicher.2017}, which implements several types of differential imaging strategies: ADI (Angular Differential Imaging), SDI (Spectral Differential Imaging), ASDI (Angular and Spectral Differential Imaging), and RSDI (Reference Star Differential Imaging). Various algorithms are offered in conjunction with the aforementioned strategies: cADI \cite[][]{Marois.2006}, LOCI \cite[][]{Lafreniere.2007}, T-LOCI \cite[][]{Marois.2010}, PCA \cite[][]{Soummer.2012}. Direct stacking and subtraction of an azimuthally averaged profile are also available. The package applies similarly to IRDIS and IFS data cubes.

 For all these algorithms the outputs follow the same rationale and include:

\begin{itemize}
\item{
A final image using one of the algorithms, in which all wavelength channels are stacked.
}
\item{
The same final image corrected from flux loss caused by ADI-induced self-subtraction of companion flux
}
\item{
A final cube, similar to the final image, but without stacking it in the wavelength dimension (with one frame per wavelength)
}
\item{
The corresponding signal to noise ratio map (one map for each wavelength)
}
\item{
Detection limits: contrast curve at 5 sigma in each wavelength as a function of the angular separation
}
\end{itemize}

%The Level 1 reduced master cubes are then used as input for several distinct Angular Differential Imaging \citep[ADI][]{Marois.2006,Lafreniere.2007} algorithms. We use the Speckle calibration (Specal) package, described by \cite{Galicher.2017} using notably TLOCI-ADI \citep{Marois.2010} routines given in \citet{Galicher.2011}.  This package also enables to perform classic ADI, Principal Component Analysis-ADI \citep{Soummer.2012} {\bf REF OK???}, Reference Differential Imaging as well as non ADI stacking, for use notably when observations are acquired in field tracking mode.
%  For all these algorithms the outputs follow the same rationale and include:
% \begin{itemize}
%\item{Several final reduced residual images, the stack of all science frames in all spectral channels after subtraction of the central star light using one the above algorithms.}
%\item{A final reduced residual image corrected from flux loss caused by ADI-induced self-subtraction of companion flux.}
%\item{A signal to noise map (one map for each wavelength).}
%\item{A final reduced residual cube, similar to the  final reduced residual image, but without stacking it in the wavelength dimension (with one frame per wavelength).}
%\item{Detection limits as a function of the angular separation and contrast curve at 5 sigma in each wavelength.}
%\end{itemize}
%

Full Level 2 data are available from the PC and a subset of images (final reduced median image, contrast curve, SNR maps) are available at the TDB.  In both cases  public data are accessible with the public login and private data with private login.
Examples of Level 2 images are shown on Fig.~\ref{fig6}.

%\begin{itemize}
%\item 	The final reduced median image in fits and png formats (data of Level 2 such as described in sect. \ref{level2})
%\item  The contrast curve at 5 sigma in each wavelength (data of Level 2)
%\item  The SNR cubes with a 2 D map for each wavelength (data of level 2)
%\item The map of detection and the json files described above and in Sect. \ref{level3} (data of level 3)
%\end{itemize}

\subsection{The DIVA archive}
\label{DIVA}

The DIVA (Direct Imaging Virtual Archive) has been built for reduced data for Adaptive Optics imaging instruments other than SPHERE 
(NaCo, HST, Keck, Gemini, Subaru, LBT) and contains astrometry and photometry of the detections as well as images. They include 
a large part of pre-SPHERE surveys. The procedure and data are described by \cite{Vigan.2017}.
Note that the DIVA and SPHERE data are in the same relational database (as described in Sect.~2.2.2); their portal are currently different but we plan to merge these two sites in the near future.
%XXXXXXX
%Note that the same data are available in the TDB and in the DIVA archive at the origin of this project and we plan to merge these two sites. 
%In practice, they already access the same database.\\

\section{Future developments}

\subsection{Overcoming our current limitations}

Though we strive to provide a very comprehensive service of SPHERE data reduction, we do not yet cover all kind of data that this very versatile instrument can provide. Our main current limitations are listed below. We are currently working to resolve each of the following issue, the time scale depending on the manpower available at SPHERE-DC to continue developing our tools, beyond data production and existing tools maintenance.
\begin{itemize}
\item{ZIMPOL: we do not yet have a fully automated pipeline for ZIMPOL and this will be implemented beginning of 2018. As of now we can only reduce ZIMPOL data for PIs on a case by case basis, using two sets of routines: i/ Part of the reduction is performed using the ESO DRH pipeline. Because of the stripe mask on the ZIMPOL detector, all recipes include a pre-processing step that re-orders the image rows into logical and viewable images. In imaging mode, the pre-processed imaging science frames are then calibrated (for each camera) by subtracting the master imaging bias and the master imaging dark, and divided by the corresponding intensity imaging flat field.  The calibrated frames are de-dithered and de-rotated and then saved as intermediate products.  The final step combines all these calibrated, de-dithered and de-rotated frames, using a standard mean algorithm, to produce one reduced image per detector.
ii/ 
For the polarisation data, the same processes are applied to the different measurement groups with regards to each Stokes parameters (Q [Q+, Q-] and/or or U [U+, U-], then homemade routines are applied to provide polarisation maps (intensity, degree of linear polarisation, polarisation angle, polarized flux).
What remain to be done is the integration of all these recipes as a workflow associated to associating rules in the PC. } 
\item{IRDIS Polarimetry: we only provide the basic reduction steps that are common between non-polarimetric and polarimetric scientific data. The reduction of polarimetric IRDIS data with procedures taking into account as much as possible instrumental effects will therefore be soon implemented.}
\item{IRDIS Long Slit Spectroscopy: we only provide the basic reduction steps that are common between imaging and spectroscopic scientific data, i.e. similar to the ESO pipeline. Further analysis remains to be defined and implemented and may depend on the scientific case. 
Our current approach is to implement the routines developed by \cite{Vigan.2016b} for the specific reduction of IRDIS data acquired in the long-slit spectroscopy mode. These routines consist in a combination of the standard DRH recipes with custom IDL routines designed to simplify the reduction and analysis of LSS data, going from the raw data to a set of clean, aligned image cubes that can be analysed with speckle-subtraction techniques adapted to LSS data \cite[e.g.][]{Vigan.2008}.}
\item{IRDIS Observations without star centering sequence or observations with saturated or otherwise invalid star centering sequence are currently only reduced to Level 1.}
\end{itemize}

\subsection{Level 3 data} 
\label{level3}

Data with more added value will be provided in the future, mostly through the TDB. 
While some Level 2 data were part of the ESO pipeline (although we propose more added value), Level 3 data are completely outside the scope of the ESO pipeline and different added value data may be produced depending on the scientific cases. 
We plan to produce the Level 3 data described below, which concerns detections associated to their property and tools to visualize them, but more will be produced in the future for other science cases. 

We recall that detection are objects identified in the field of view of coronagraphic images when searching for exoplanets: they can be either bounded companions, background stars. 
Data of Level 3 would then be all informations related to the detections found in such Level 2 images, in particular their
astrometric and photometric characterization. These data can already be produced by the SpeCalcharac \cite[][]{Galicher.2017} algorithm included in the DC.
The algorithm however requests a visual identification of candidate companions and therefore cannot be applied in an automatic workflow yet. We are developping automated detection algorithms compatible with SpeCalcharac, but their current ratio of true positive over false positive detections is still less reliable than human analysis. 
The TDB is already able to deal with these data of Level 3:
%%%%%for the private SHINE survey. Thus, the TDB is already able to deal with these data of level 3. They are currently transferred into the TDB via the json files described in Annexe \ref{tables}.
The tools developed at the TDB allows the users to determine whether the detections are background objects or possible companions. 
It is also possible to use forms in the TDB to search and sort by criteria (ex: stars that have bound objects, etc.).

\begin{figure}[ht!]
 \centering
 \includegraphics[width=1.0\textwidth,clip]{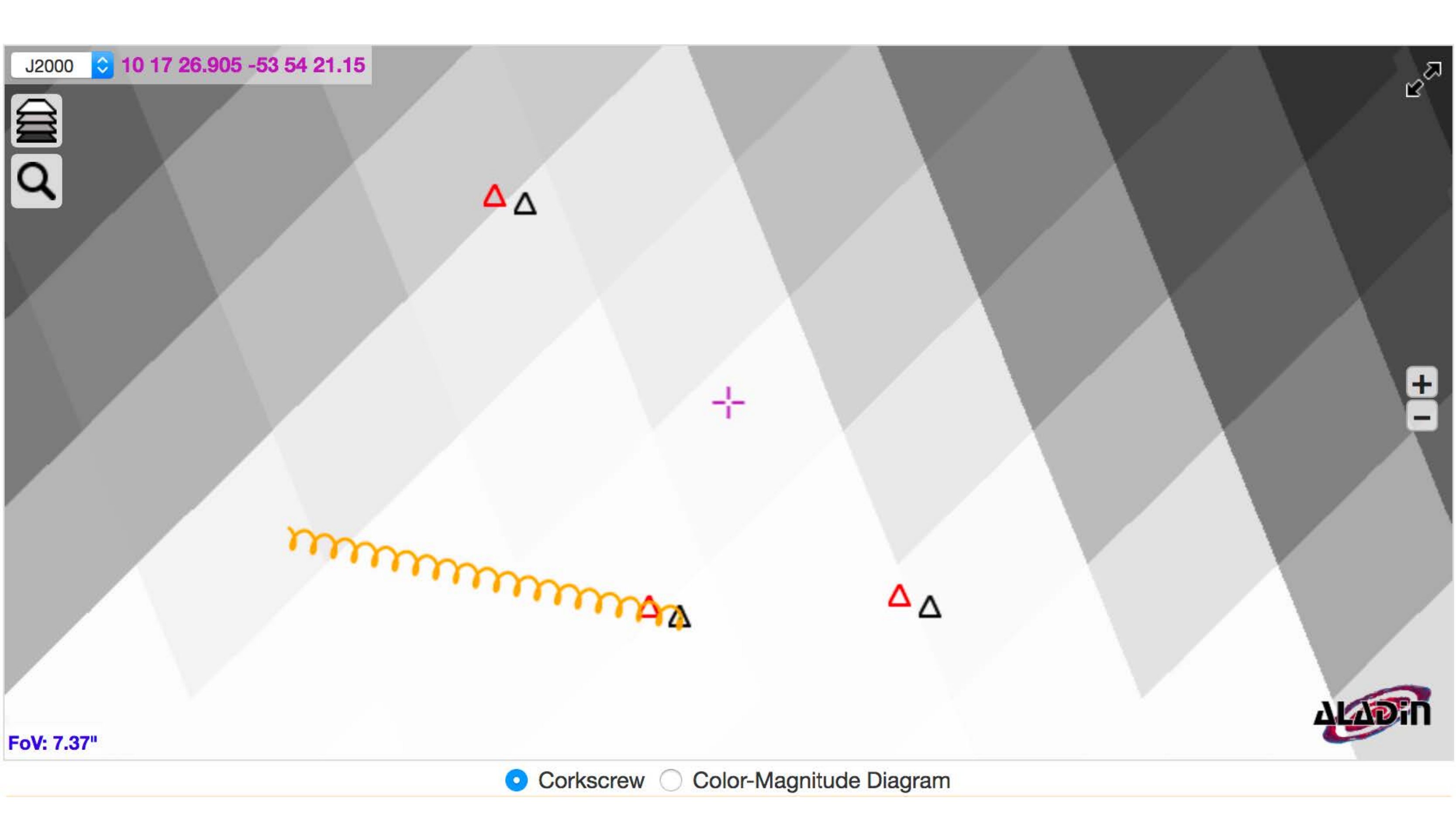}
%%% Note the ABSENCE of the extension .pdf  !
  \caption{Three detections around a star of the DIVA project (NACO and NICMOS instruments) at two different dates (dark and red). The astrometry curve of the background stars is shown in yellow.}
  \label{fig_astro}
\end{figure}

To help the users to determine the nature of the detections (background stars or bound companions), we have therefore developed several dynamics charts using Aladin tools and javascript libraries such as 
plotly.js\footnote{https://plot.ly/javascript/} and chart.js\footnote{http://www.chartjs.org/}. For example, it is possible to show one detection, 
or all the detections found around one star, or all the detections in the TDB, in color magnitude diagrams: MH2[mH2- mH3], MK1[mk1-mk2], etc., which are crucial to determine their nature. 
Reference points of known cold objects are super-imposed on these graphics.
An other tool of astrometry allows to visualize all the detections at all the epochs of observation around a selected star (as illustrated for example in Fig. \ref{fig_astro}), 
and to superimpose the astrometric curve of the background stars. An admin tool allows to regroup the detections observed at several dates so that the users can see graphically 
the detections nature (i.e. put in a same group) around one star (background stars, bound objects, etc.).
the TDB shows all its power, by plotting all together, on the same graphic, data that come from any surveys (NaCO, SPHERE, Keck, Hubble, etc.).

Future work will therefore include an automatisation of the detection procedure and the develoment of other graphics, and statistical tools to provide this Level 3 data.

\subsection{New algorithms for Level 2 data and frame sorting}

We plan to regularly add state-of-art differential imaging algorithms to our toolbox to follow the fast development of such routines in the high-contrast imaging community. Beyond striving to provide such up-to-date algorithms, we also want to explore the unique opportunities opened by coupling a direct imaging reduction center to a massive database taking advantage of Reference star Differential Imaging while avoiding self-subtraction inherent to ADI, SDI or ASDI. Using a large library of coronagraphic images will allow to perform in a more efficient way as it was done recently for HST NICMS \citep{Choquet.2016}.

Another issue concerns frame selection, i.e. the elimination of low quality images before going from Level 1 to Level 2 (and above) data: which frames in a given science sequence were affected by poor AO correction ? This is currently investigated using science data cubes and wavefront sensors data (see Sect.~5.4) and will lead to complementary Level 2 data after frame selection, in addition to information on the data quality. 
 
\subsection{Exploitation of observatory metadata}

%We have started to implement some instrument monitoring routines, that introduce into the database the atmospheric conditions informations as well as the wavefront sensor data (coherence time, wind speed, seeing, photons per lenslet, etc) with the densest time sampling available (typically of 30s). These are crucial to interpret the reduced data. In the near future we will couple this information with the detection limits as measured on the simultaneous final reduced science data. This will open new opportunities in various fields:
%\begin{itemize}
%\item{ Instrument monitoring: e.g. are performances stable over time in quantitatively similar conditions?}
%\item{ Observations optimisation : e.g what is the the minimum number of photons per lenslet necessary to achieve a given contrast for a given set of observing conditions ?}
%\item{ Data reduction with PSF libraries : which observations in the database have similar AO correction as the science sequence considered, and could be used as optimal PSF for differential imaging ?}
%\end{itemize}
%

Apart from science detector data, the system archives some contemporary statistics of the wavefront sensors data. This provides direct information on the outer turbulence conditions and the achieved AO correction quality. Temporal sampling for these statistical data is typically every 30s up to now, and on-going discussions with ESO currently push to increase this sampling rate to every 10s.

On a statistical manner, such information is useful to monitor the instrument performance and also to help and predict what performance to expect according to conditions. Such an analysis motivates some current change at ESO to modify the user-specified constraints to execute given OBs (ESO Observing Blocks) in service mode according to turbulence speed rather than seeing, which appear much more relevant to guarantee the desired performance level \citep{Milli.2017arXiv}.

\begin{figure}[ht!]
 \centering
 \includegraphics[width=1.3\textwidth,clip]{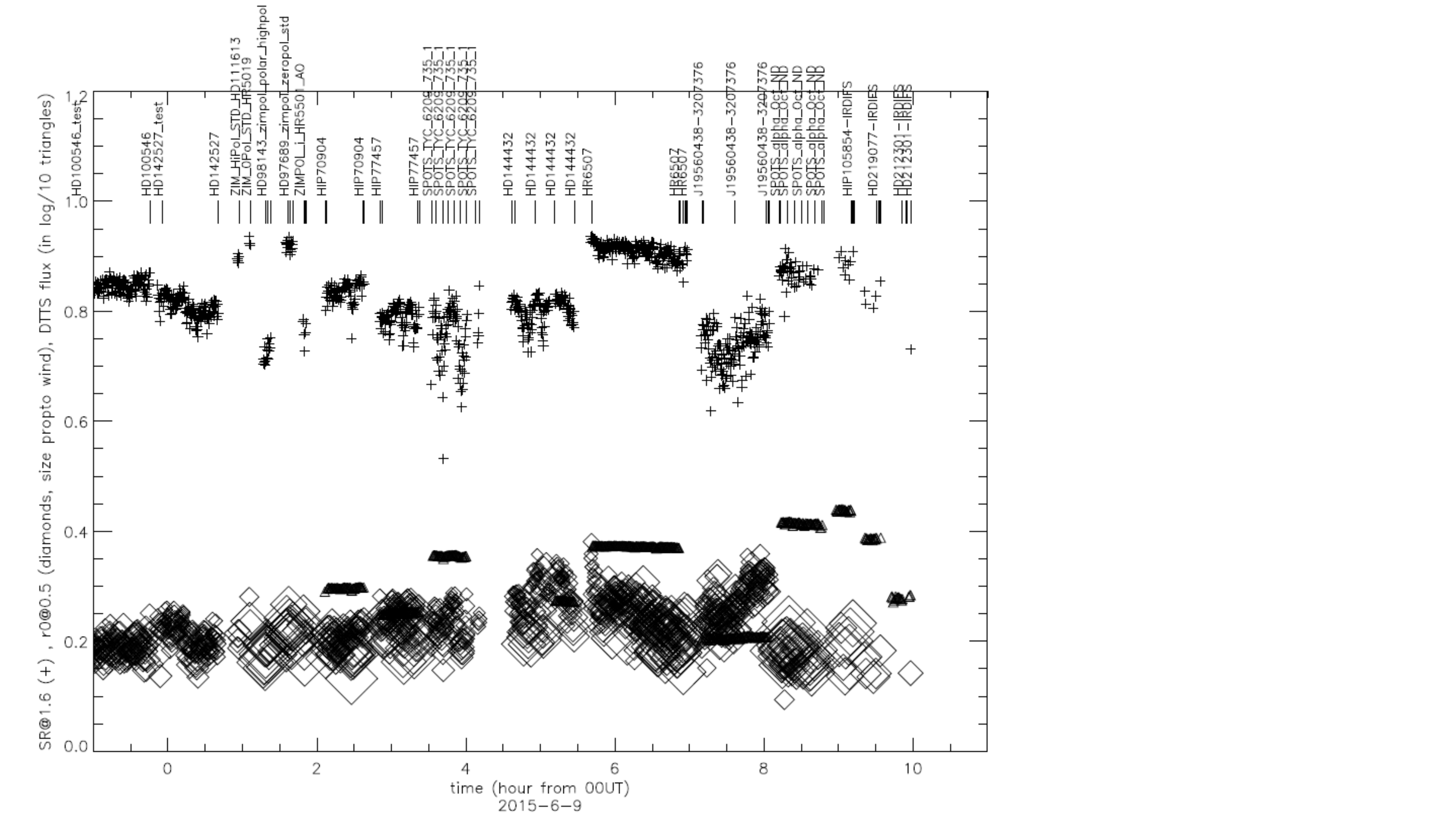}     
%%% Note the ABSENCE of the extension .pdf  !
  \caption{Example of the night conditions overview as extracted from AO data. It includes an estimate of the achieve AO image quality (Strehl ratio at 1.6 mic imaging wavelength), spatial and temporal scale of the turbulence (r$\_0$ and $t\_0$).}
  \label{OA1}
\end{figure}

\begin{figure}[ht!]
 \centering
 \includegraphics[width=1.\textwidth,clip]{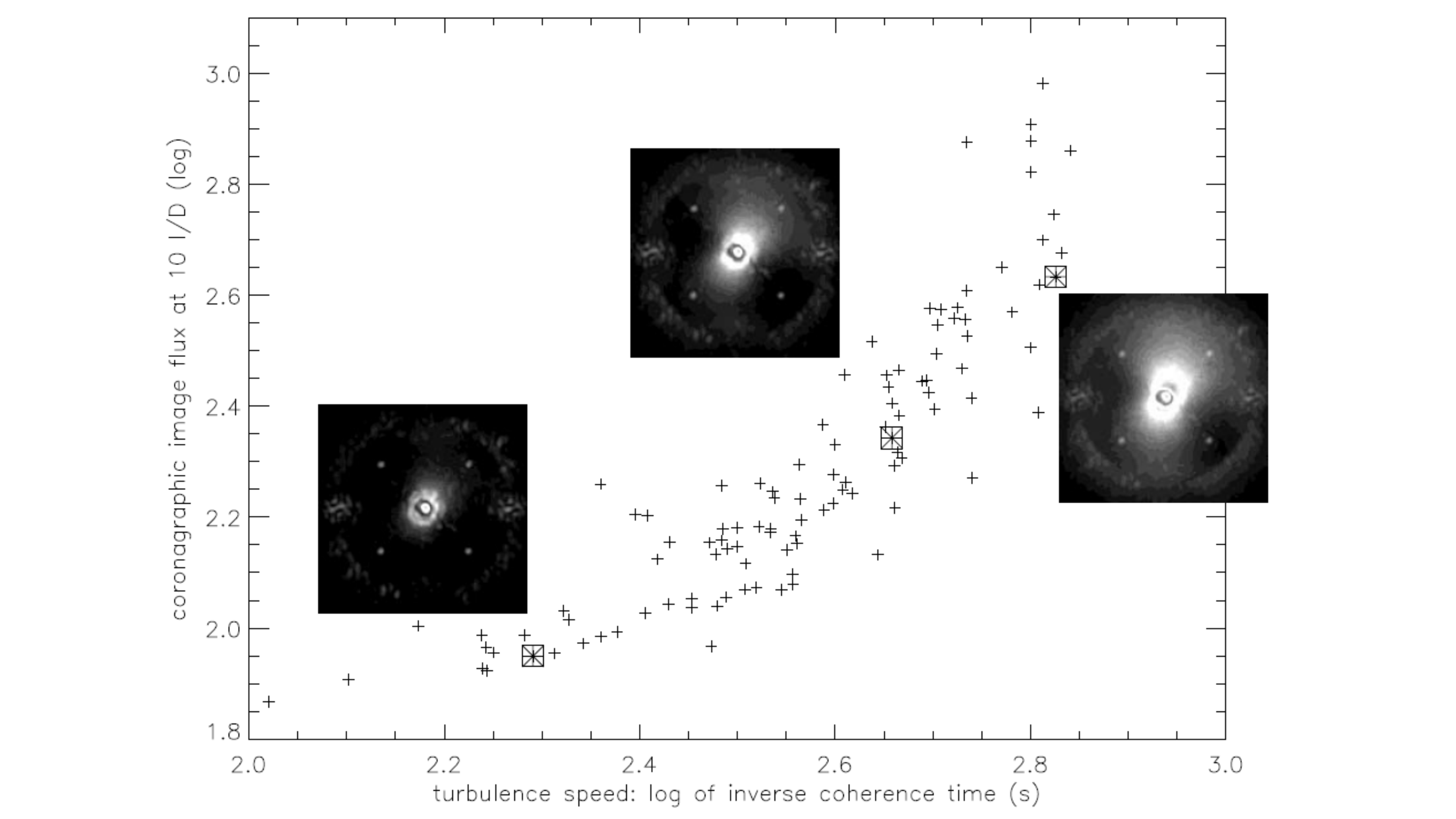}     
%%% Note the ABSENCE of the extension .pdf  !
  \caption{Correlation between the coronagraphic image intensity over an annulus at 10 $\lambda$/D vs the AO sensor estimate of the turbulence speed (log of the inverse coherence time expressed in s), along a 1-hour sequence on a bright object, in variable conditions. The turbulence speed estimator is a very good proxy for the coronagraphic flux leaks.
}
  \label{OA2}
\end{figure}

On an a posteriori data reduction perspective, we have started development to extract relevant information and associate then directly to the science data product with the following interests and goals:

\begin{itemize}
\item{
{\it Night and/or program overviews}: the AO data provide a nice overview of the conditions and image quality obtained during a night or a program, which is much more relevant than the generic ESO seeing monitor. It includes both the seeing and turbulence speed and the achieved image quality correponding to the science data (Fig.~\ref{OA1}).
}
\item{
{\it Data selection for a given data set reduction, or building classes of PSF libraries}: The coronagraphic images by construction hide the stellar PSF core and direct indication of the AO quality SR is lost. The AO sensor data provide this in real time. Also, the turbulence speed is directly correlated to the coronagraphic flux leaks (especially in the direction of the wind), as illustrated in Fig.~\ref{OA2}. The AO data are thus very valuable to guide the image selection among data cube that are often heterogeneous. These AO data are currently being associated to reduced science frames. It will be used to support the optimal selection of frames to obtain the best contrast of a given data set; it will also guide the construction and classification of large PSF libraries according to conditions and support the choice of the best PSF among this library to enhance the power of « reference star differential imaging » \citep[RSDI][]{Gerard.2016,Wahhaj.2016}.
}
\end{itemize}

\subsection{Data format and tools}

The data available at the TDB will soon be also available in a HCI-FITS format \cite[][]{Choquet.2014,Hagan.2017} as aready done for the DIVA project\footnote{http://cesam.lam.fr/diva/index/format} (see Sect.~4.4).

Other tools may also be available in the future, for example tools to prepare survey observations. For example, the coordinators of the SHINE survey can prepare their observation catalogues directly from the TDB. This option, in discussion, could be developed for other surveys and PIs but might require to be adapted to deal with a lot more instrumental configurations.

\subsection{Virtual Observatory layer}

We plan to develop an interoperability layer for both databases, with a Virtual Observatory (VO) access to the public data. 
This will allow direct requests on the database. The typical fields will be : coordinates, target name, instrument (to be completed). 
This layer will be implemented for both PC and TDB databases.
On a longer term basis, it may be possible to implement our catalogue at the CDS: this would allow easier cross-identification with other 
catalogues at CDS.

\section{Conclusion}

The SPHERE Data Center has been operational for more than two years, with success. It allowed to provide reduced data with a good reactivity to many observers, and the first public reduced data have been made recently available through the PC and TDB interface (corresponding to a major subset of ESO P95 period, i.e. April-September 2015), for Levels 1 and 2. The SPHERE Data Center is gathering a strong expertise on SPHERE data and is in a very good position to propose new reduced data in the future, as well as improved reduction procedures. Some tools to extend the coverage and provide new added value are currently being developed, and some of the functionalities are almost ready (for example concerning the Level 3 data "detections", functionality already existing at the TDB). 
This position opens interesting perspectives for future instruments (for example high contrast instruments on the  E-ELT).

%\noindent Recently, more than 90 public observations of the SPHERE P95 period have been imported in the TDB (data of level 2). 
%The private SHINE observations (and their reductions, detections) are also automatically imported in the TDB. 
%The management (importation and visualization) of level 3 data  (see Sect.~5.2; detections with their astrometric and photometry informations) is already possible in the TDB.
%But, at this time, only SHINE private data of level 3 have been imported. They will become public at the end of the GTO.\\

%%%{\bf METTRE LA PHRASE DE REMERCIEMENT ICI}

% -------------------------
\begin{acknowledgements}
Acknowledgement: We thank the colleagues from the SPHERE consortium who have helped us with feedback on our services and by providing some tools, special thanks to Janis Hagelberg.
%, (COMPLETER ? PASSER EN CO-AUTEUR ? VERIFIER QU'ON N'OUBLIE PERSONNE ; METTRE AUSSI DANS LE TEXTE ? XXXX).
The SPHERE Data Center received support from the LabEx OSUG@2020 (Investissement d'avenir - ANR10 LABX56). We thank the OSUG-DC and CIMENT teams and in particular Fran\c{c}oise Roch, Bruno Bzeznik, Bernard Boutherin, Jean-Charles Augereau. We used the jMCS framework developed by the JMMC and available at https://github.com/JMMC\_OpenDev/jMCS. We thank Gilles Duvert for his help in using GDL. This work has made use of the SIMBAD database and ALADIN tools operated at the CDS, Strasbourg, France. 
We thank the CeSAM for its support. We thank the computer department at IPAG for its support as well as the administrative support both at IPAG and OSUG. 
This project has received funding from the European Union’s Horizon 2020 research and innovation programme under the Marie Skłodowska-Curie Grant agreement No. 665501 with the research Foundation Flanders (FWO) ([PEGASUS] 2 Marie Curie fellowship 12U2717N awarded to MM.
\end{acknowledgements}

%% The following lines are required when using BibTEX (strongly encouraged!):
\bibliographystyle{aa}  % A&A bibliography style file (aa.bst)
\bibliography{biball} % your references in file: Yourfile.bib

\appendix
\section{Description of the tables in the Target Data Base}
\label{tables}

The main tables of the data base are described below (not exhaustive).
We find this hierarchical logic (Observations $->$ Reductions $->$ Detections ) in all the web site and notably in the json files
used to import the reduced products from any surveys. 
\\

\noindent \textbf{Observation:}\\

\noindent observation date (JD)
instrument name \\
telescope name\\
survey name\\
target name (simbad)\\
true north (deg)\\
true north error (deg)\
filter name	\\
parallactic angle (deg)\\
pixel scale on the CCD (mas/pixel)\\
pixel scale error on the CCD (mas/pixel)\\
observing setup (generic name for all the config. Ex: IRDIFS)\\
unique observation id\\

\noindent \textbf{Reductions:}\\

\noindent validation flag\\
software name used for the reduction	\\
version	 of the software\\
algorithm used (PCA, TLOCI, etc.)\\
Date of the reduction (JD)\\
Address of the reduction on the DC\\
Unique id number of reduction\\

\noindent \textbf{Detections:}\\

\noindent \textbf{Detection 0:}\\
wavelength (micron)\\
magnitude\\
magnitude error\\	
signal to noise\\
x position (mas)\\	
x position error (mas)\\
y position (mas)\\
y position error (mas)\\	
separation (mas)\\
separation error (mas)\\	
position angle (deg from North)\\
position angle error (deg)\\
\noindent \textbf{Detection 1:}\\
etc.\\

\end{document}